\documentclass[10pt]{IEEEtran}
\usepackage{graphicx}
\usepackage{tikz}
\usepackage{amsmath}
\usepackage{times}
\usepackage{subfigure}
\usepackage{caption} 
\usepackage{algorithm}
\usepackage{algorithmic}
\usepackage{tabularx}
\usepackage{xspace}
\usepackage{booktabs}
\usepackage{multirow}
\usepackage{enumitem}
\usepackage{makecell}
\setlist{nosep}
\usepackage{url}

\usepackage{amsfonts}
\usepackage{url}
\usepackage{makecell}
\usepackage{orcidlink}

\hypersetup{
  colorlinks=false,
  linkbordercolor=white,
 urlbordercolor=white,
pdfborder={0 0 0}
}

\newcommand{\SecureSpike}{{\textsc{\small{PrivSpike}}}\xspace}

\begin{document}

\title{PrivSpike: Employing Homomorphic Encryption for Private Inference of  Deep Spiking Neural Networks}

\author{
    Nges Brian Njungle\orcidlink{0009-0006-3393-6851}$^{1}$,
    Eric Jahns\orcidlink{0009-0004-5511-7975}$^{1}$,
    Milan Stojkov\orcidlink{0000-0002-0602-0606}$^{2}$,
    and Michel A. Kinsy\orcidlink{0000-0002-1432-6939}$^{1}$\\
    $^{1}$STAM Center, Ira A. Fulton Schools of Engineering, Arizona State University, Tempe, AZ 85281, USA\\
    $^{2}$Faculty of Technical Sciences, University of Novi Sad, Novi Sad, Serbia\\
    Emails: nnjungle@asu.edu, jjahns@asu.edu, stojkovm@uns.ac.rs, mkinsy@asu.edu
}

\maketitle

\begin{abstract}
Deep learning has become a cornerstone of modern machine learning. It relies heavily on vast datasets and significant computational resources for high performance. This data often contains sensitive information, making privacy a major concern in deep learning. 
Spiking Neural Networks (SNNs) have emerged as an energy-efficient alternative to conventional deep learning approaches. Nevertheless, SNNs still depend on large volumes of data, inheriting all the privacy challenges of deep learning.
Homomorphic encryption addresses this challenge by allowing computations to be performed on encrypted data, ensuring data confidentiality throughout the entire processing pipeline.

In this paper, we introduce \SecureSpike, a privacy-preserving inference framework for SNNs using the CKKS homomorphic encryption scheme. \SecureSpike supports arbitrary depth SNNs and introduces two key algorithms for evaluating the Leaky Integrate-and-Fire activation function: (1) a polynomial approximation algorithm designed for high-performance SNN inference, and (2) a novel scheme-switching algorithm that optimizes precision at a higher computational cost.
We evaluate \SecureSpike on MNIST, CIFAR-10, Neuromorphic MNIST, and CIFAR-10 DVS using models from LeNet-5 and ResNet-19 architectures, achieving encrypted inference accuracies of 98.10\%, 79.3\%, 98.1\%, and 66.0\%, respectively.
On a consumer-grade CPU, SNN LeNet-5 models achieved inference times of 28 seconds on MNIST and 212 seconds on Neuromorphic MNIST. 
For SNN ResNet-19 models, inference took 784 seconds on CIFAR-10 and 1846 seconds on CIFAR-10 DVS.
These results establish \SecureSpike as a viable and efficient solution for secure SNN inference, bridging the gap between energy-efficient deep neural networks and strong cryptographic privacy guarantees while outperforming prior encrypted SNN solutions.
\end{abstract}


%
%
%

\section{Introduction}
\label{sec:intro}

In recent years, machine learning has revolutionized various industries by driving significant technological advancements. Its applications span diverse fields, including medical diagnosis, traffic prediction, image recognition, speech recognition, product recommendation, self-driving cars, virtual personal assistants, and natural language processing \cite{ml_applications}. Central to these advancements are deep neural networks, particularly convolutional neural networks (CNNs), which leverage large-scale datasets to achieve high precision, especially in domain-specific tasks \cite{khan_sok}. 
Despite their effectiveness, CNNs require substantial energy in both training and inference.
These limitations have sparked great interest in Spiking Neural Networks (SNNs) as a promising alternative \cite{patrick_fpga}.

Unlike conventional deep neural network approaches that synchronously process static frames, SNNs operate on asynchronous, event-driven principles that more closely mimic biological neural systems. This key distinction enables SNNs to process spatiotemporal information with remarkable energy efficiency and responsiveness~\cite{ClaudioCimarelli-2025}.
Utilizing these dynamics, Lenz et al.~\cite{GregorLenz-2024} demonstrated that SNNs deployed on Intel’s Loihi 2 neuromorphic hardware~\cite{IntelLoihi2} can reduce the energy consumption of standard CNN inference by up to $246\times$ on MNIST.
These results underscore the potential of SNNs in low-power computing scenarios where traditional deep learning models are impractical. 
Another major application domain of SNNs is in event-based data processing.  
SNNs have shown considerable promise in processing time-sensitive and event-based biomedical and sensory data, where traditional machine learning approaches fall short. This is because the temporal dynamics of these datasets naturally fit into the event-driven data processing mechanisms of SNNs. For instance, Xu et al.~\cite{xu} demonstrated superior electromyographic pattern recognition ability using SNNs,  outperforming cyclic CNNs of the same architecture by up to 50\%. 
Sola et al.~\cite{xu2023novel} also reported state-of-the-art results on gesture recognition from an event-based dataset using SNNs with up to $25$\% better performance over recurrent neural networks. These examples highlight the superiority of SNNs in handling data with intrinsic temporal structure, making them ideal for applications in neuromorphic computing and energy-constrained settings.
However, similar to other deep neural network approaches, SNNs also require significant data to train, thus inheriting the critical data and model privacy issues of deep neural networks. 
These concerns are particularly apparent when models are inferred on highly sensitive private information like biomedical and sensory data.

Although privacy-preserving techniques such as homomorphic encryption \cite{9936637}, multi-party computation \cite{zhang}, differential privacy \cite{Gong}, and trusted execution environments \cite{Li2023ASO} have been extensively explored for privacy-preserving computations in CNNs, little attention has been paid to data and model privacy issues in SNNs \cite{njungle2025guardianml}.
Among these techniques, homomorphic encryption (HE) has emerged as the most suitable for privacy-preserving neural networks due to its strong cryptographic security guarantees and efficient communication properties despite the higher computational overhead it brings. 
It enables computations directly on encrypted data, thus ensuring data confidentiality throughout the computational pipeline. 
Among HE schemes, the Cheon-Kim-Kim-Song (CKKS) scheme is particularly well suited for neural network applications due to its support for approximate arithmetic and parallel data processing \cite{ckks}. 
Despite its advantages, utilizing CKKS for SNNs presents distinct challenges. Firstly, the efficient evaluation of non-linear functions, such as the leaky integrate-and-fire (LIF) activation function, is difficult since the scheme does not inherently support the evaluation of non-linear operations.  
Secondly, SNNs encode data using temporal and spatial dynamics into multiple time-steps, further increasing the computational and memory requirements of encrypted inference. 
Previous works applying HE to SNNs, such as \cite{farzad}, have attempted to address these issues by proposing distinct privacy-preserving SNN inference solutions; however, all previous solutions have fallen short in terms of performance and scope of evaluation. 

In this work, we introduce \SecureSpike, a comprehensive and efficient framework for privacy-preserving SNN inference using the CKKS scheme. 
\SecureSpike tackles critical challenges in privacy-preserving SNN inference by introducing two novel algorithms for the widely-used LIF activation function, alongside optimized implementations of all SNN layers. These two LIF algorithms offer a trade-off between performance and accuracy of models. 
The first algorithm employs polynomial approximation to handle the non-linear components of the LIF, while the second utilizes a scheme-switching technique to evaluate spikes of encrypted inputs under the TFHE scheme securely \cite{tfhe}. To the best of our knowledge, this is the first work to propose such algorithms for encrypted-domain LIF evaluation. 
Additionally, we incorporate architectural optimizations that significantly reduce the computational overhead and memory requirements of evaluating encrypted data on SNNs.
We evaluated our work using the MNIST, CIFAR-10, Neuromorphic MNIST (N-MNIST), and CIFAR-10 DVS datasets. While MNIST and CIFAR-10 are popular deep neural network datasets, N-MNIST and CIFAR-10 DVS are popular neuromorphic datasets mainly used with SNNs. We infer these datasets using state-of-the-art SNN models from the LeNet-5 and ResNet-19 architectures. 
Our evaluation demonstrates that \SecureSpike scales efficiently, with models of identical architecture using almost the same amount of memory, independent of the number of time steps.  This consistent resource utilization simplifies deployment and enhances scalability across different model types and datasets.
In terms of inference latency, we observe that the inference time of SNNs grows linearly with the increase in the number of time steps, under the same HE security parameters. 
This increase in latency is attributed to the temporal dynamics inherent to SNNs, which involve iterative updates over multiple time steps. Nonetheless, \SecureSpike’s efficient scaling makes this additional cost manageable.
Furthermore, the accuracy of privacy-preserving SNN models closely matches that of their plaintext counterparts, demonstrating the practicality of our implementations and methods. 
Notably, our scheme-switching LIF algorithm maintained a higher precision, with encrypted inference results deviating by $0.8\%$ from plaintext outputs in the LeNet-5 model using MNIST.
On the LeNet-5 architecture, shows about 34$\times$ and 50$\times$ better inference time compared to related works.
In this work, we make the following concrete contributions:
\begin{itemize}
    \item We propose \SecureSpike, an efficient, open-source, privacy-preserving SNN inference framework using the CKKS homomorphic encryption scheme. 
    \item We introduce and implement two algorithms for the LIF activation function tailored for high performance and precision in encrypted SNN computations. The first algorithm utilizes Chebyshev polynomials to approximate the non-linear component of the LIF activation function, while the second employs the scheme-switching technique to compute high-precision results of this component using the TFHE scheme. 
    \item We evaluate \SecureSpike using four widely studied datasets and two well-established SNN architectures using eight SNN models.  Specifically, we conduct experiments on the MNIST and CIFAR-10 datasets, as well as on the neuromorphic N-MNIST and CIFAR-10 DVS datasets. The N-MNIST  and CIFAR-10 DVS ensure the relevance of our work on neuromorphic and event-based data processing, while the MNIST and CIFAR-10 datasets show the compatibility of \SecureSpike with traditional ML workloads.
\end{itemize}

\section{Related Works}
\label{sec:related_works}

In 2022 and 2023, Casaburi and Nikfam et al. introduced the first HE framework for SNNs \cite{farzad}, \cite{farzad_thesis}. Their approach employed the Brakerski/Fan-Vercauteren (BFV) HE scheme \cite{bfv} to perform encrypted inference on SNNs. Experiments were conducted on the FashionMNIST dataset using LeNet-5 and AlexNet architectures. The framework required approximately 930 seconds to infer a single encrypted image using the SNN LeNet-5 model. For the SNN AlexNet model, the authors estimated an inference time of 901,800 seconds per encrypted image. Despite the high latency, the approach achieved a notable accuracy of 96.5\% for encrypted inference on the LeNet-5 model.
The framework also introduced a significant communication overhead due to the computation of the LIF activation functions on the client side. Their setup requires the server to transfer encrypted data to the client for the evaluation of activation functions. The client decrypts the data, computes the activation function in plaintext, re-encrypts the results, and sends them back to the server to continue the inference process. 
The authors employed a Tesla P100-PCIE GPU, an Intel(R) Xeon(R) Gold 6134 @ 3.20GHz CPU, and 100GB of RAM.
Although they showcased the feasibility of privacy-preserving SNNs, the framework remains inadequate for real-world adoption due to the substantial computational resources required, the communication overhead it introduces, and the high inference latency of the models. 

Still in 2023, Li et al. introduced FHE-DiCSNN \cite{Li2023EfficientPC}. Their work leveraged the discrete nature of the Fully Homomorphic Encryption over the Torus (TFHE) scheme \cite{tfhe} to perform privacy-preserving inference on SNNs. They chose the TFHE scheme because it naturally aligns with the discrete spike characteristic of SNNs. 
Their work was evaluated on the MNIST dataset and achieved an accuracy of 95.10\% on encrypted data.  FHE-DiCSNN took 0.75 seconds to infer a single encrypted MNIST image. 
The primary limitation of  FHE-DiCSNN is that its evaluation is done on a small network comprising only 30 neurons.
In practical SNN research, LeNet-5 is the most commonly used shallow architecture, consisting of roughly 60,000 neurons.
Hence, approximating the total inference time of the framework on an SNN LeNet-5 model gives an inference time of 1,500 seconds per encrypted image.
The TFHE scheme aligns naturally with SNNs, but it is significantly slower in deep learning tasks compared to schemes like CKKS. This limitation arises from its lack of support for parallel data processing, a capability widely leveraged to accelerate computations on encrypted data~\cite{ckks}.

In contrast to the aforementioned works, our approach utilizes the CKKS scheme to provide privacy-preserving inference in SNNs. We leverage optimizations, such as parallel data computation for high high performance evaluation of all linear layers. 
We propose two distinct algorithms for evaluating the LIF activation function in the encrypted domain. 
The first algorithm employs the Chebyshev polynomial approximation of the LIF function.
The second algorithm utilizes scheme switching, allowing for high precision evaluation of the non-linear component of the LIF in TFHE. 
This hybrid approach leverages the strengths of both schemes, thus using CKKS for efficient linear computation and TFHE for high-precision non-linear evaluation. 
We infer our models on encrypted data, and notably, our SNN LeNet-5 model, which employs the approximation method for LIF, takes just 28 seconds to infer an encrypted MNIST image on a consumer CPU. 
Additionally, we evaluated SNN models based on the ResNet-19 architecture, further demonstrating great performance, scalability, and efficient resource utilization for encrypted data processing within \SecureSpike.
This comprehensive evaluation with models from widely used SNN architectures and diverse datasets (standard and event-based)  highlights the practical applications of \SecureSpike.

\section{Background}
\label{sec:background}

\subsection{Homomorphic Encryption}
\begin{figure}[http]
    \begin{center}
    \includegraphics[width=\linewidth]{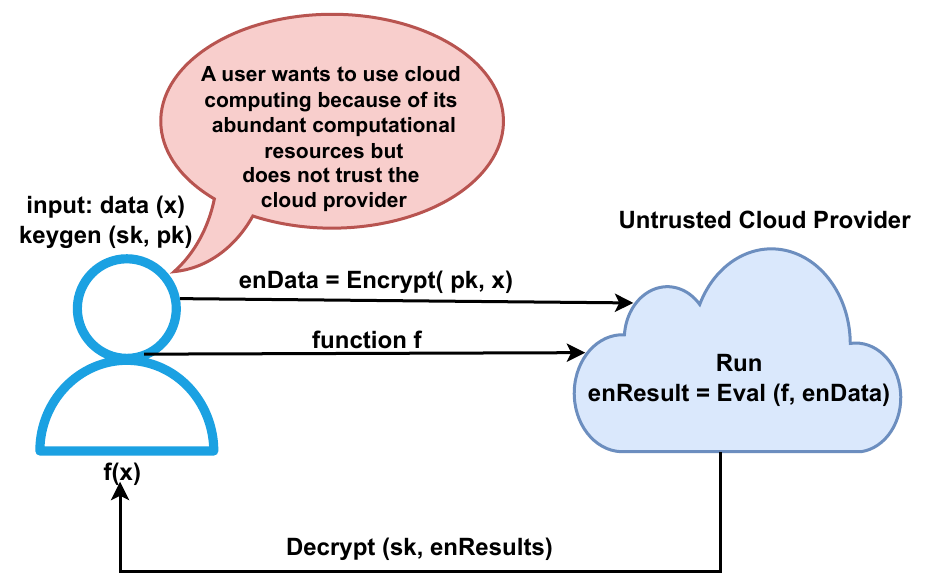}
    \captionsetup{justification=centering}
    \caption{ A HE outsourced cloud computation scenario showing a user who encrypts data and evaluates a function in the cloud. The encrypted results are sent back to the user, who decrypts them to get the evaluated results.}
    \label{fig:fheworks}
    \end{center}
\end{figure}

Homomorphic Encryption (HE) is an advanced cryptographic protocol that allows computation on encrypted data.
The term “homomorphic” also means “same form”. 
In mathematical terms, objects have the same form in plaintext as in ciphertext if there exist equivalent mathematical operations in both the plaintext and ciphertext domains that produce corresponding outputs in their respective domains.
Figure \ref{fig:fheworks} illustrates how HE works in an outsourced computational scenario. 
Given a user wants to outsource computation but does not trust the cloud provider, HE gives the user the ability to encrypt and store data in a secure form in the untrusted cloud server and then evaluate functions securely on this data without leaking any information. 

Since the demonstration of Fully Homomorphic Encryption (FHE)  by Gentry \cite{gentry}, the field has seen significant research breakthroughs and the introduction of numerous FHE schemes \cite{fhe_vul}. 
CKKS \cite{ckks}, BFV \cite{bfv}, BGV \cite{bgv}, and TFHE \cite{tfhe} make up state-of-the-art HE schemes, each optimized for different application scenarios. 
For instance, TFHE processes bitwise circuits, while BFV and BGV handle integer computations. In contrast, CKKS is designed for floating-point arithmetic computations, making it well-suited for this work.

\subsubsection{Cheon-Kim-Kim-Song (CKKS) scheme}
This scheme was released in 2016 by Cheon, Kim, Kim, and Song, and it is the only mainstream HE scheme today that operates on Approximate Arithmetic Numbers \cite{ckks}. 
The scheme anchors its security on the security assumptions of the non-probabilistic hard problem of the ring version of the Learning with Errors problem \cite{ringlwe}. 
Mathematically, let $\mathbb{Z}$, $\mathbb{R}$, and $\mathbb{C}$ represent Integers, Real numbers, and Complex numbers, respectively. If given a vector $b \in \mathbb{Z}_q^m$ and matrix \(A \in \mathbb{Z}_q^{m\times n}\), the standard LWE problem is looking for an unknown vector \(s \in \mathbb{Z}_q^n \) such that:
\begin{equation}
As + e = b \;mod\;q\
\end{equation}
where $e$ is an error vector introduced from random samples through an error distribution, and $q$ is a large prime number.

The Ring Learning with Errors (RLWE) problem extends the LWE problem to polynomial rings over finite fields. This ring structure enables CKKS efficient parallel data representation and processing. 
The primary homomorphic operations defined in the CKKS scheme include addition, multiplication, rotation, and bootstrapping. 
Bootstrapping allows for the refreshing of noise in the ciphertext. 
The Rotation operation shifts the data encoded in the ciphertext to the right or left, while multiplication and addition allow for the multiplication and addition of ciphertext messages, respectively.
In 2018, Cheon et al. released an optimized version of CKKS on the residue number system \cite{rns_ckks} that transforms the CKKS ciphertext to Double Chinese-Remainder-Transform polynomials, allowing for fast multiplication using the Number Theoretical Transform (NTT), which reduces the complexity of multiplications from $O(n^2)$  in basic modular polynomial multiplication to $O(n \log n)$.

A key feature of CKKS is its ability to perform ciphertext slot packing, wherein multiple pieces of data are encoded into the coefficients of a single large polynomial.
This packing mechanism facilitates a Single Instruction Multiple Data (SIMD) computational paradigm, where a single homomorphic operation can process multiple values in parallel. 
This approach significantly accelerates the runtime of CKKS by enabling the simultaneous evaluation of multiple data points, making the scheme highly efficient for applications that require large amounts of data through batching. 
 Let $R = \mathbb{Z}[X] / (X^N + 1)$ represent a ring of polynomials modulo $X^{N} + 1$, where $N$ is a power of 2. The CKKS scheme converts a message $v \in \mathbb{C}^{N/2}$ to $R$. 
Using the SIMD approach, homomorphic operations can be performed in parallel across all encoded slots, effectively processing multiple data elements in the ciphertexts simultaneously, thus reducing the cost of operations in such settings. However, at most half the slots in a ciphertext can be independently used due to the need to balance complex conjugate pairs in the ciphertext encoding process.  
This constraint arises because CKKS encodes plaintext values in the complex domain, and thus, decryption requires the conjugate symmetry of the encoded coefficients. 
The use of approximate arithmetic in CKKS, combined with its support for the operations above and SIMD parallel processing capabilities, makes it particularly well-suited for large-scale data processing tasks, such as those involved in SNNs.

\begin{figure*}[ht!]
    \centering
    \vspace{-0.1in}
    \includegraphics[width=\linewidth]{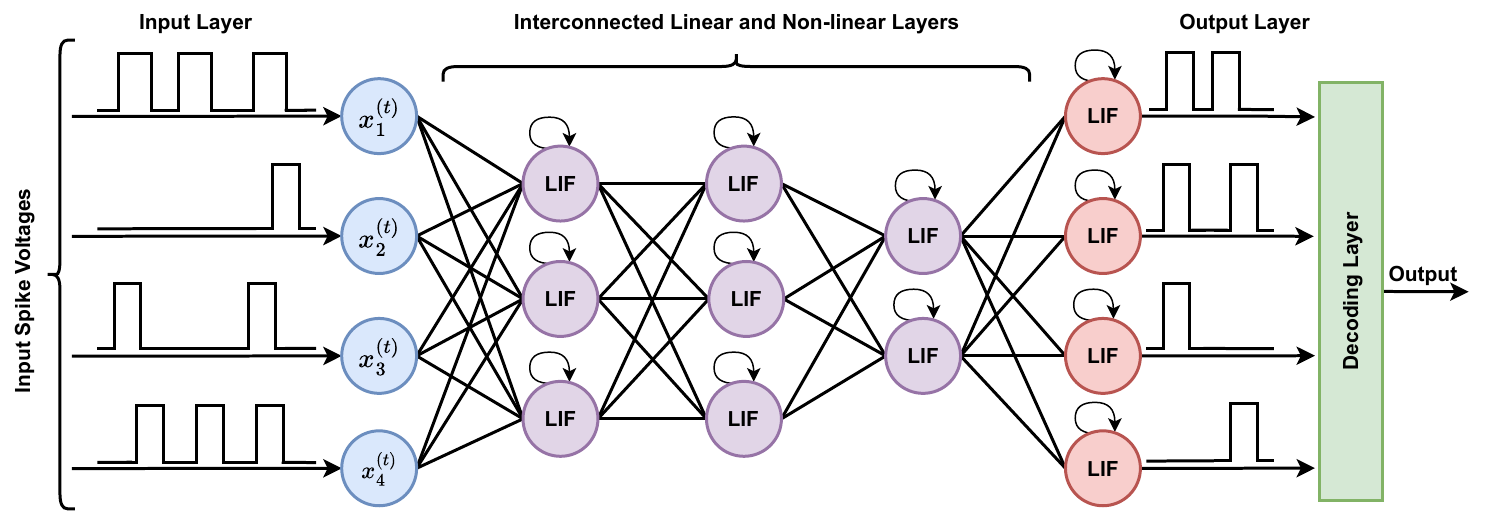}
    \captionsetup{justification=centering}
    \caption{Overview of a spiking neural network. Spikes are input across discrete time steps and propagated through linear and non-linear layers. Linear layers extract spatial features, while non-linear spiking dynamics capture temporal dependencies via recurrent processing. The output layer accumulates spikes over time and decodes them into a prediction.}
    \label{fig:snn_arch}
\end{figure*}

Open-source HE Libraries, such as Microsoft SEAL \cite{sealcrypto}, HElib \cite{helib}, TFHE-rs \cite{TFHErs}, and OpenFHE \cite{OpenFHE}, provide diverse implementations of HE schemes~\cite{njungle2025safety}. These libraries facilitate the adoption of HE by abstracting the complexities of developing HE schemes. 
Thus, these libraries are crucial for promoting the practical use of HE in real-world applications and research.
Among the available CKKS implementations, OpenFHE offers the most advanced support for the scheme. Its implementation supports optimizations such as SIMD, bootstrapping, scheme switching to TFHE, and efficient rotation strategies, which is why we chose it for this work.

\subsection{Spiking Neural Networks (SNNs)}
SNNs are a biologically inspired class of artificial neural networks that process information through discrete spike events, closely mimicking the communication mechanisms of neurons in the brain \cite{ghosh2009spiking}. 
Unlike traditional neural networks, which utilize continuous-valued activations, SNNs operate over multiple discrete time steps, dynamically and sparsely processing input data \cite{marchisio2020spiking}. 
This temporal and event-driven paradigm allows SNNs to inherently encode and process time-varying information, making them particularly energy-efficient and well-suited for neuromorphic hardware such as Intel Loihi, IBM TrueNorth, and BrainScaleS \cite{spikinghard}.
The core computational units of SNNs are spiking neurons, which simulate the dynamics of biological neurons. 
For a neuron \(i\), the membrane potential \(V_i(t)\) evolves (\(t\)) in response to incoming spikes, and a spike is emitted when \(V_i(t)\) crosses a predefined threshold. This spike resets the membrane potential, introducing temporal dynamics to the computation. These dynamics enable SNNs to process and encode spatiotemporal patterns natively \cite{han2021survey}. 
At each time step, only a subset of neurons and synapses are active, resulting in a significant reduction in computational overhead compared to traditional networks that process all features simultaneously. This property aligns well with hardware constraints in low-power and real-time scenarios such as event-based cameras.

Figure~\ref{fig:snn_arch} illustrates the structure and temporal operation of an SNN. Input data is encoded as spike events and processed over multiple discrete timesteps. As spikes propagate through the network, linear layers perform spatial transformations, and spiking neurons integrate inputs over time to extract temporal features, maintaining a membrane potential that enables them to retain information between time steps. The network’s output layer aggregates spiking activity across all time-steps to decode a final prediction, effectively capturing the dynamic nature of time-dependent data.

\section{Secure Spiking Neural Network Layers}
\label{sec:snnlayers}

This section presents a detailed overview of the layers and processes involved in the design and development of SNNs. For each layer, we describe how it is adapted for encrypted computation within the \SecureSpike framework.

\subsection{Input Layer}
The input layer of an SNN is the initial stage where raw input signals are processed and converted into spike-based representations suitable for subsequent computations~\cite{ho2021tcl}. The layer forms the foundation, ensuring that the temporal and spatial characteristics of inputs are preserved and appropriately formatted for processing through the network. As illustrated in Figure~\ref{fig:snn_arch}, the input layer accepts voltage signals corresponding to input spike voltages, which encode the temporal dynamics of the data. These input spike voltages are represented across multiple dimensions, denoted by indices $i$, which define the spatial and temporal resolution of the input~\cite{almomani2019comparative}, \cite{yang2023spiking}. The layer converts these input signals into spiking activity, a process that generally involves SNN input encoding schemes such as rate coding \cite{kim2022rate} and temporal coding \cite{mostafa2017supervised}. 
Its primary function is to bridge the gap between conventional input data and the event-driven nature of SNNs.

To avoid the high latency, computational cost, and performance degradation introduced by encoders, the first convolutional layer typically replaces a dedicated encoder in deep SNNs when processing static data, as it can perform feature extraction and spike transformation simultaneously. Unlike traditional encoders, which are explicitly designed to preprocess and convert input signals into spike trains, a convolutional layer can learn spatial and temporal patterns from raw input data while generating spike-based activations. This dual functionality eliminates the need for a separate encoding mechanism, streamlining the architecture and reducing computational overhead \cite{encoders}. 
This approach is adopted in \SecureSpike as it offers several advantages in SNN computations, especially when combined with HE and high-dimensional inputs. It simplifies the data preprocessing pipeline while maintaining flexibility and robustness by enabling the networks to adaptively encode and process data without relying on hand-crafted encoding rules. 
Additionally, this design reduces the number of time steps required to achieve the desired accuracy thus improve temporal efficiency.

\subsection{Convolution Layer} 

In deep learning, the convolution operation is a transformation to extract features from input data \cite{namatevs2017deep}.
The convolution layer applies a set of filters, called kernels, to the input data by sliding the filters over the data, applying element-wise multiplication, and summing the results to produce a single output \cite{Kamath2019}, \cite{JIANG2022102954}. This process results in a feature map that highlights specific features the filter detects. 
Mathematically, the convolution operation is defined as follows: Let the input tensor be denoted by \( X \), with a shape of \( (d, h, w) \), where:
 \( d \) is the depth (or number of channels),
 \( h \) is the height,
and \( w \) is the width of the input matrix.
Let the kernel \( K \) have a shape of \( (c_{\text{out}}, c_{\text{in}}, k_h, k_w) \), where:
\( c_{\text{out}} \) is the number of output channels,
 \( c_{\text{in}} \) is the number of input channels,
 \( k_h \) is the height of the kernel,
 and \( k_w \) is the width of the kernel.
The convolution operation computes the output feature map \( Y \) with a shape of \( (c_{\text{out}}, h_{\text{out}}, w_{\text{out}}) \), where \( h_{\text{out}} \) and \( w_{\text{out}} \) are the height and width of the output, respectively. The convolution operation can be expressed mathematically as:
\begin{align}
    Y_{c_{\text{out}}, h_{\text{out}}, w_{\text{out}}} = \sum_{c_{\text{in}}=1}^{c_{\text{in}}} \sum_{i=1}^{k_h} \sum_{j=1}^{k_w} X_{c_{\text{in}}, h_{\text{out}} + i - 1, w_{\text{out}} + j - 1} \cdot K_{c_{\text{out}}, c_{\text{in}}, i, j}
\end{align}

\( X_{c_{\text{in}}, h_{\text{out}} + i - 1, w_{\text{out}} + j - 1} \) represents the element of the input tensor at the corresponding depth, height, and width location and \( K_{c_{\text{out}}, c_{\text{in}}, i, j} \) represents the value of the kernel at the corresponding output channel, input channel, and kernel location.
The sum is taken over all input channels \( c_{\text{in}} \) and over the height and width of the kernel \( k_h \) and \( k_w \).
Employing the convolution operation na\"ively with HE requires considerable storage and computational resources. A more efficient approach to the convolution operation for HE was introduced by Juvekar et al. \cite{gazelle}, called vector encoding, which utilizes SIMD optimization in HE to perform multiple convolution operations in parallel. This approach has also been adopted for use in many privacy-preserving CNN works, such as \cite{kim_fhe}, \cite{lee22e}, \cite{lowmemory20}.

To use this approach, we generalize it and implement a configurable convolutional layer reusable across all forms of networks. This layer accepts an encrypted input along with metadata specifying the input width, number of input and output channels, and kernel size.
The input to the first convolutional layer is a flattened and encrypted spatial tensor, represented as a SIMD ciphertext. 
This ciphertext is formed by packing the input values channel-by-channel into a single vector, which is then encrypted. Each subsequent layer in the network processes an encrypted input, which is the ciphertext output from the preceding layer.

To perform the encrypted convolution, we generate $k^2 - 1$ rotations of the input SIMD ciphertext (where $k$ is the kernel size). Each rotated ciphertext corresponds to a shift in the input and is multiplied by a precomputed SIMD vector containing repeated copies of the corresponding kernel weight. 
The unrotated input is similarly multiplied by a SIMD vector containing the repeated first kernel element. 
All resulting ciphertexts from these multiplications are summed to produce a final ciphertext encoding the full convolution output. 
Finally, to obtain a compact convolution, we apply $w_{\text{out}}$ ciphertext rotations to drop intermediate error values introduced during by the encrypted computation.
Figures~\ref{fig:con_rots} and~\ref{fig:con_sum} in the Appendix illustrate this convolution process for a $2 \times 2$ kernel.

\subsection{Average Pooling}
The pooling layer in deep learning is primarily used for dimensionality reduction \cite{gholamalinezhad2020pooling}. It is conceptually similar to a convolution layer, but differs because its filters are uniform. The two widely used pooling layers in deep learning are \textit{Maximum Pooling} and \textit{Average Pooling}. Maximum Pooling selects the maximum value from each region covered by the pooling kernel. At the same time, Average Pooling computes the average of all values within the region covered by the pooling kernel \cite{PASSRICHA20195}.
Pooling layers often use striding, allowing the filter to skip certain input regions. This effectively reduces the data size used in subsequent network layers.
In the context of CKKS, \textit{Maximum Pooling} is computationally expensive as it involves evaluating a non-linear greater-than function. In contrast, \textit{Average Pooling} is more efficient and well-suited for encrypted computations as it can be performed using a single homomorphic multiplication and  $k^2-1$ homomorphic additions. Average pooling is mathematically defined as follows:
\begin{align}
\text{Output}(i, j) = \frac{1}{k^2} \sum_{x=0}^{k-1} \sum_{y=0}^{k-1} \text{Input}(i+x, j+y),
\end{align}
This operation involves summing all kernel-mapped elements and multiplying by a pre-computed $\frac{1}{k^2}$. 
Our average pooling implementation uses the same algorithm as in the convolution layer to generate the different rotations of the input tensor. We then sum the rotated ciphertexts and multiply with a precomputed value, $\frac{1}{k^2}$. 
Figure \ref{fig:pooling} in the Appendix show average pooling for $k=2$. 

\subsection{Fully Connected Layer}

Fully connected layers map all input data to an output space by learning a set of weights that define the relationships between all input and output neurons in the neural networks \cite{intro_cnn}. In this layer, every input neuron is connected to every output neuron, thus allowing the network to learn complex global patterns. Mathematically, if given an input vector \( x \) of size \( n \), weights \( W \) of size \( m \times n \), biases \( b \) of size \( m \), and output vector \( y \) of size \( m \), the computation for the fully connected layer can be expressed as:
\begin{align}
    y_k = \sum_{i=1}^{n} W_{ki} x_i + b_k \quad \text{for each } k = 1, 2, \dots, m
\end{align}
With SIMD ciphertext packing in CKKS, operations in the fully connected layer become highly efficient, requiring only one multiplication and \( n \) additions for each output neuron, making the fully connected layer the most computationally efficient component in \SecureSpike. 
For each output neuron in any fully connected layer, a single SIMD multiplication is performed for \( W_i \times x_i \), where \( W \) is the weights vector of the output neuron, \( x \) represents the input neurons vector, and \( i \) denotes the index of the corresponding input neuron. Following this, \( n \) homomorphic additions are carried out to compute the result for the output neuron. These computations are repeated \( n_{\text{out}} \) times for all output neurons in the layer.
To aggregate the results, \( n_{\text{out}} \) rotations are applied to combine the output neuron values into a single ciphertext representing the layer’s outputs. \( n_{\text{out}} \) rotations also means \( n_{\text{out}} \) rotation keys for all corresponding indices.

\subsection{Leaky-Integrate and Fire (LIF) Layer}
Activation Functions are essential components of deep neural networks, enabling them to learn complex, non-linear relationships within data \cite{DUBEY202292}. In SNNs, they play a unique and critical role by determining how neurons process and transmit information through discrete spikes, mimicking the behavior of biological neurons. Unlike traditional neural networks, which use activation functions that output continuous values, SNNs employ spike-based mechanisms to encode information temporally, making them ideal for tasks that require dynamic and time-sensitive processing.
Several activation mechanisms are commonly used in SNNs, including but not limited to the Integrate-and-Fire (IF) model, Leaky Integrate-and-Fire (LIF) model, and the Hodgkin-Huxley model \cite{nunes2022spiking}. These activation functions enable temporal information processing by incorporating dynamics such as integrating input signals over time, voltage decay, and spike generation. Among these functions, the LIF model is the most widely adopted due to the balance it provide between computational efficiency and biological realism. 

In the LIF model, neurons accumulate input currents over time  $I(t)$ into a state variable known as the membrane potential, denoted as \( V(t) \) \cite{teeter2018generalized}. The neuron emits a spike when the membrane potential crosses a predefined threshold \( Th \). After firing, the neuron undergoes a reset, returning \( V(t) \) to a resting state \( V_{\text{rest}} \). The membrane potential dynamics are governed by a differential equation that incorporates a decay term, reflecting potential dissipation over time. This decay is controlled by the membrane time constant \( \tau_m \), which influences the rate at which the voltage returns to its resting state without input. The behavior of the LIF neuron is expressed as shown in Equation \ref{eq:lif_ex}.
\begin{equation}
\label{eq:lif_ex}
    \tau_m \frac{dV}{dt} = -(V - V_{\text{rest}}) + I
\end{equation}
A neuron emits a spike when \( V \geq Th \), where \( Th \) is the threshold voltage. Upon firing, the membrane potential is reset to \( V_{\text{rest}} \), and the neuron is optionally subjected to a refractory period, during which it temporarily becomes inactive. 
To further simplify this equation for implementation purposes, the discrete version of the equation is expressed as shown in Equation \ref{eq:discr}.
\begin{equation}
    V(t+1) = \left(1 - \frac{\Delta t}{\tau_m}\right)V(t) + I(t) \label{eq:discr}
\end{equation}

Algorithmically, we define the  LIF model as shown in Algorithm \ref{alg:lif_algo}. It contains a non-linear comparison operation, used to determine whether the membrane potential exceeds the threshold. However, this operation cannot be directly evaluated in the CKKS scheme since the scheme does not support the evaluation of non-linear functions. 
To address this limitation, we propose two distinct versions of this algorithm. We employ the Chebyshev polynomials to approximate the comparison function in a computationally feasible manner as well as apply scheme-switching for high precision evaluation within the TFHE scheme.

\begin{algorithm}
\caption{The Leaky Integrate-and-Fire Algorithm}
\label{alg:lif_algo}
\begin{algorithmic}[1]
\STATE \textbf{Input:}, $V(t)$, $I(t)$,  $t$
\STATE \textbf{Initialization:}
\STATE $V_{rest} \gets 0$
\STATE  $\tau_m \gets 0.25 $
\STATE  $\text{Th} \gets 0.05$
\STATE $\text{spikeValue} \gets 0$ 
\FORALL{time steps $t = 1, 2, \dots, T$}
    \STATE \textbf{If initial timestep:}
    \IF{$t = 1$}
        \STATE $V(t) \gets I(t)$
    \ELSE
        \STATE $V(t) \gets \tau_m V(t-1) + I(t)$
    \ENDIF

    \STATE \textbf{Spike Decision:}
    \IF{$V(t) > \text{Th}$}
        \STATE $\text{spikeValue} \gets 1$
        \STATE $V(t) \gets V_{\text{rest}}$ 
    \ELSE
        \STATE $\text{spikeValue} \gets 0$ 
    \ENDIF
    \STATE $V(t) \gets (1 - \text{spikeValue}) \cdot V(t)$
\ENDFOR
\end{algorithmic}
\end{algorithm}

\subsubsection{Polynomial Approximation of the LIF Function}
Chebyshev polynomials are a set of orthogonal polynomials defined over the interval $[-1, 1]$, and they are particularly useful for approximating non-linear functions with high precision \cite{gil2012non}.
This approach has been extensively used in privacy-preserving CNN works with HE as the state-of-the-art approach for evaluating the ReLU function~\cite{lowmemory20}

Using this approach, we then convert the comparison function present in LIF and express it as a series of polynomials, which becomes CKKS-friendly and computationally feasible.
In the context of the LIF neuron model, we experimentally determine efficient values to scale $V(t)$ to $[-1, 1]$ for each dataset and each layer and set them as an input parameter for the function since the values of $V(t)$ are the encrypted results of the previous convolution operation and are not within the desired range for evaluation.  
We replace the comparison operation $V(t) > Th$ with an approximation function based on Chebyshev polynomial evaluation of the $>$ operation. 
With Chebyshev polynomial approximation, the degree of the polynomial expansion $N$ determines the accuracy of the approximation of the spiking activation function. The higher the degree, the better the approximation to the non-linear function. On the downside, the higher the polynomial degree, the higher the computational complexity of the approximation function. 
With further application of SIMD, we design an FHE-friendly algorithm for the  LIF function based on Chebyshev approximation as presented in  Algorithm \ref{alg:fhe_lif_algo}. 

\begin{algorithm}
\caption{Secure LIF Algorithm Using Chebyshev Approximation}
\label{alg:fhe_lif_algo}
\begin{algorithmic}[1]
\STATE \textbf{Input:}, $V(t) \gets \textbf{vector}$, $I(t) \gets \textbf{vector}$,  $t$, $scaleValue$
\STATE \textbf{Output:}, $Spikes(t) \gets \textbf{vector}$,  $V(t) \gets \textbf{vector}$
\STATE \textbf{Initialization:}
\STATE  $\tau_m \gets 0.25 $
\STATE  $\text{Th} \gets 0.5/scaleValue$
\STATE  $\text{N} \gets 50$
\STATE \textbf{If initial timestep:}
    \IF{$t = 1$}
        \STATE $V(t) \gets I(t)$ 
    \ELSE
        \STATE $V(t) \gets \tau_m V(t-1) + I(t)$
    \ENDIF
    \STATE $V(t) \gets \text{ScalingFunction}(V(t), scaleValue, [-1, 1])$
    \STATE \textbf{Spike Decision:}
    \STATE $spikes(t) \gets \text{ApproximateGreaterThan}(V(t), \text{Th}, $N$)$
    \STATE $V(t) \gets (1 - Spikes(t)) \cdot V(t)$
\end{algorithmic}
\end{algorithm}

This algorithm takes an input membrane potential and accumulated voltage ciphertexts, and scales their sum using predetermined scaling value to fit all values of the ciphertext within [-1, 1]. 
We then apply SIMD friendly approximation of the comparison function over a scaled threshold. 
To calculate the accumulated voltage of the next timestep, we subtract the result the spikes multiplied by the input voltage from one and return both the spikes and accumulated voltage. 
The spikes are then propagated to the next layer of the network while the accumulated voltage is added to input current of the next time step. 
Experimentally, we ran the Chebyshev polynomial Approximation function with different values of $N$ and determined that $N = 50$ is an appropriate default polynomial degree for the approximation of the spiking function as it efficiently balances the computational complexity and accuracy. In our evaluation, we used this default value for all models to establish a uniform baseline for comparison. A Chebyshev polynomial of this degree requires seven multiplications for evaluation in CKKS, which is reasonable given the CKKS parameters required for secure deep SNN models. We also parameterize the value of $N$ in our implementation, as different values of $N$ can produce slightly different levels of precision for the function.

\subsubsection{Scheme-Switching Evaluation of LIF Function}

HE schemes offer different trade-offs in terms of efficiency and computational capabilities. These differences make certain schemes more suitable for specific types of encrypted operations. The CKKS scheme is particularly effective for approximate arithmetic on real-valued data. It supports SIMD computations, which allows for parallel evaluation of linear operations over large encrypted datasets.

In contrast, schemes such as FHEW~\cite{fhew} and TFHE~\cite{tfhe} are designed for high-precision computation based on Boolean logic. These schemes are well-suited for evaluating non-linear functions like comparisons, sign determination, and the floor function. 
To evaluate a function in these schemes, the function must be expressed as a Boolean circuit, which is built from basic logic gates such as AND, OR, XOR, and NOR gates. This gate-level representation enables accurate computation of discrete functions within the constraints of HE and is defined as the basic operations of these schemes.
Although FHEW and TFHE achieve high accuracy through Boolean computations, they do not support SIMD parallelism. As a result, they are computationally expensive when used for large-scale data processing as well as evaluating complex deep circuits like those found in SNNs.

To leverage the strengths of TFHE and CKKS in \SecureSpike, we also propose a hybrid LIF algorithm for high precision evaluation of the LIF function based on scheme-switching. 
This algorithm combines both schemes for data processing in SNNs.
All linear components in privacy-preserving SNNs, such as convolutional and fully connected layers, are computed using CKKS to leverage its parallel processing capabilities. 
For high precision in the non-linear components of the LIF, the TFHE scheme is employed for the comparison operation while CKKS is maintained for all other linear operations found in the algorithm.
The idea of switching between encryption schemes was introduced by Christina et al. in the CHIMERA framework~\cite{chimera}.
Scheme switching is possible between CKKS and TFHE because both schemes are based on the LWE problem. 
This shared mathematical foundation enables ciphertexts to be converted between these schemes while preserving the encrypted values and semantic meaning, without any compromise in the security and privacy of the encrypted data.

Our LIF scheme-switching algorithm proposed and implemented in \SecureSpike is shown in Algorithm \ref{alg:fhe_lif_algo_switch}. It accepts three ciphertexts: one for the encrypted membrane potential, the encrypted accumulated voltage, and the encrypted threshold vector.
A threshold vector, filled with the LIF model's standard threshold value of $0.5$, is pre-computed and encrypted using the CKKS scheme for secure use in  SNNs.
In the LIF model, both the encrypted membrane potential and the threshold are converted into TFHE ciphertexts. A Boolean comparison circuit is then used to determine whether the neuron should spike by comparing the membrane potential and the threshold value.
The result is a binary ciphertext that contains 0s and 1s, indicating the inverse spike states of all neurons. 
The spikes are calculated by subtracting all resulting SIMD binary ciphertext from 1 ensuring that all spiked neurons are 1s while 0s are non-spike neurons. 

\begin{algorithm}
\caption{LIF Algorithm using the Scheme-switching Technique}
\label{alg:fhe_lif_algo_switch}
\begin{algorithmic}[1]
\STATE \textbf{Input:}, $V(t) \gets \textbf{vector}$, $I(t) \gets \textbf{vector}$, $T(t) \gets \textbf{vector}$ $t$, \text{vectorSize}
\STATE \textbf{Output:}, $Spikes(t) \gets \textbf{vector}$,  $V(t) \gets \textbf{vector}$
\STATE \textbf{Initialization:}
\STATE  $\tau_m \gets 0.25 $
\STATE \textbf{If initial timestep:}
    \IF{$t = 1$}
        \STATE $V(t) \gets I(t)$ 
    \ELSE
        \STATE $V(t) \gets \tau_m V(t-1) + I(t)$
    \ENDIF
    \STATE \textbf{Spike Decision:}
        
    \STATE Set slots of $c_{\text{enc}}$ to $\text{vectorSize}$
    \STATE $tc_{\text{fst}} \gets \text{CKKSSwitchToTFHEW}(V(t), \text{vectorSize})$
    \STATE $tc_{\text{sec}} \gets \text{CKKSSwitchToTFHEW}(T(t),\text{vectorSize})$
    \STATE $cResult(t)_{} \gets \text{TFHECompare}
    (tc_{\text{fst}}, tc_{\text{sec}}, \text{vectorSize})$
     \STATE $spikes(t) \gets 1 - \text{cResults(t)}$
    \STATE $V(t) \gets cResults(t) \cdot V(t)$
\end{algorithmic}
\end{algorithm}

\subsection{Output Decoding Layer}
The Output Decoding Layer in SNNs is responsible for interpreting the spiking activity of neurons and mapping them into meaningful outputs suitable for specific tasks, such as classification or regression \cite{yang2023spiking}. 
Unlike traditional neural networks, where output neurons encode information through continuous activations, SNNs utilize discrete spike events to represent data. 
The decoding layer translates these spike-based representations into a usable format by aggregating spike counts, analyzing spike timings, or employing temporal coding strategies to extract relevant information from the spike trains generated by the network.

In \SecureSpike, the Output Decoding Layer aggregates the network's outputs over multiple steps to produce a final prediction. At each time step, the spike-based outputs are summed, effectively integrating temporal information and smoothing fluctuations in spike activity. 
This decoding approach combines the advantages of temporal dynamics with a straightforward summation mechanism, enabling robust and simple handling of the sparse and discrete nature of spike-based representations. The network can refine its predictions by integrating information across time, compensating for noise or variability in individual time steps. 
This method aligns well with our approach, as summing multiple output ciphertext is a computationally efficient operation.
The aggregated results can be decrypted to obtain the inference results. The predicted label is determined as the class corresponding to the maximum value in the aggregated ciphertext.

\section{Threat Model}
Our threat model aligns with that of previous works based on the security guarantees of HE. Specifically, we adopt a semi-honest adversarial threat model, where both the client and server follow the protocol without deviation but may attempt to infer or extract private information from encrypted data.
Security in \SecureSpike is assured through the CKKS  scheme with implementation offered by OpenFHE. 
These primitives allow us to build SNN layers and futher models which allow  computations on encrypted data without revealing plaintext values.
The client encrypts input data before sending it to the server, allowing SNN inference on \SecureSpike models to be performed on encrypted data. 
After inference, the server returns the encrypted result to the client, who can decrypt it using their private key. 
We also assume that the client is the sole holder of the private key, thus the only party capable of decrypting the inference output of the model.
We assume the attacker to be any entity seeking to gain information about the user's private data. This includes the model owner, the computation server, and any intermediaries on the communication channels.
Our threat model excludes malicious actors who intentionally perform incorrect computations, as well as scenarios in which the user's private key is compromised.

\section{Experiment}
\label{sec:experiment}
This work was evaluated on two systems: the  Ryzen 5900x CPU with 64GB of RAM (consumer-grade CPU) and 64-core AMD EPYC Milan 7713 CPUs with 256GB of RAM. 
All models that used the LIF approximation were evaluated on the Ryzen 5900X CPU, along with the LeNet-5 model that employed the scheme-switching approach for evaluating LIF. 
The ResNet-19 models that employ the scheme-switching approach were evaluated on the AMD EPYC system due to the higher memory requirements needed for scheme-switching. 
\SecureSpike is developed using FHEON as the base to provide basic neural network layers and utility functions. 
\SecureSpike is made openly available on github as an open-source project on \url{https://github.com/stamcenter/privespike}

\subsection{Spiking Neural Network Architectures}
We validate \SecureSpike using two well-known SNN architectures: LeNet-5 and ResNet-19. These architectures serve as established benchmarks in SNN research, showcasing the effectiveness of our methods and their ability to scale across multiple architectures. 
Table \ref{tab:arch_com} of the Appendix shows the details of the neural network layers, along with their input channels, output channels, kernel size, striding, and padding configurations for the models from these architectures used in this study.
The input channels of the first convolution layer are $2$ for both N-MNIST and CIFAR-10 DVS. 
The first fully connected layer of our LeNet-5 model has 256 or 576 input channels for the MNIST and N-MNIST datasets, respectively.
LeNet-5, with its relatively simple design, serves as a baseline, while ResNet-19, with its deeper and more complex architectures, allows us to test \SecureSpike against more challenging tasks. 
This contrast ensures comprehensive validation of our methods across varying architectural complexities, demonstrating the versatility of \SecureSpike for broader applications.

\subsubsection{Architectural Optimization of SNNs for HE}
Achieving efficient inference of SNNs under HE requires careful architectural design for optimal performance. Optimizations must reduce latency and computational overhead while maintaining accuracy and security guarantees. 
In this work, we adopt several key SNN architectural innovations tailored for efficiency under HE constraints.

\paragraph{Preprocessing Rotation Keys.}
We introduce an offline stage for the precomputation and generation of evaluation keys for HE-friendly SNNs. 
Here, we analyze the input tensor shapes, kernel dimensions, and channel configurations across all layers to determine the full set of ciphertext rotation indices required during inference. 
We precompute and generate the rotation indices and keys for the entire network. These keys are imported into the network and reused throughout inference, avoiding the redundancy and memory overhead associated with regenerating or reloading duplicate keys. 
This ensures consistent memory usage across inferences, regardless of the number of time steps processed.

\paragraph{Layer-wise Iterative Evaluation.}
Unlike typical plaintext SNN implementations, which process all layers sequentially across each time step, we instead evaluate each layer iteratively across all time steps before proceeding to the next layer. 
This approach is motivated by the temporal structure of SNNs, where the output at a given time step depends on the current input and the accumulated membrane potential from previous steps. Iterative layer-wise evaluation enhances data locality and throughput, aligning more closely with the resource limitations of encrypted computation.
This optimization enables the dynamic loading and unloading of model weights for individual layers during inference, significantly reducing memory requirements. However, when the model is used to infer multiple images in sequence, frequent loading and offloading of weights becomes a trade-off, as it slightly increases inference latency.

\paragraph{Bootstrapping in Deep Networks.}
To enable deep networks with LIF approximation, \SecureSpike performs bootstrapping on the membrane potential ciphertext  \( V(t) \) before computing the spiking function approximation, particularly when the timestep is greater than one. This step is essential for maintaining ciphertext integrity because the Chebyshev polynomial used for approximation, such as a degree-50 polynomial, requires a multiplicative depth of at least seven. 
Furthermore, when an average pooling layer follows immediately after a bootstrapped non-linear layer, an additional bootstrapping is applied before pooling to ensure the correctness of encrypted computations. 
In contrast, fewer bootstrapping operations are required in networks that use scheme switching to evaluate the LIF. 
In these networks, bootstrapping is applied at specific intervals, calculated with care, to minimize computational overhead.

\subsubsection{Training}

All models were trained in PyTorch, and their weights were exported as CSV files and imported into the C++ implementations of our models. 
All models were trained using 80\% of the dataset while 20\% was used as the testing set to validate the model accuracy.

\subsection{Datasets}
The MNIST dataset is a widely used benchmark in machine learning, consisting of 60,000 grayscale images of handwritten digits, each with dimensions \(28 \times 28\) pixels \cite{deng2012mnist}. To evaluate the inference accuracy of the LeNet-5 models, we use a sample of 1,000 images from the MNIST validation set.
The N-MNIST dataset extends the original MNIST into an event-based format designed explicitly for SNNs and neuromorphic computing applications~\cite{nmnist}. In this conversion, static handwritten digit images are transformed into sequences of asynchronous spike events, which are stored in AER (Address-Event Representation) format. 
To prepare this data for use in our models, we considered the structure of our ciphertext for the most efficient transformation into the encrypted domain. Each digit sample is converted into a sequence of $5$ time steps, where each time step represents a frame composed of accumulated spike events. 
Every frame contains two channels for ON and OFF polarities, formatted as a tensor of dimensions \(2 \times 36 \times 36\), which captures both the spatial and temporal dynamics of the event. 
Unlike the default \(2 \times 34 \times 34\) representation with 10 time steps commonly used in most SNN studies, our input transformation offers a more efficient alternative for HE-friendly models. This transformation reduced inference time in the encrypted domain by half while preserving plaintext accuracy. Specifically, training a LeNet-5 with the default N-MNIST transformation yielded a plaintext accuracy of 98.8\%, whereas the same dataset using our transformed inputs achieved a slightly higher accuracy of 99.02\% in the plaintext domain.
For the evaluation of \SecureSpike, we use 150 encrypted samples from the validation set of the N-MNIST dataset to evaluate our privacy-preserving SNN LeNet-5 models.

The CIFAR-10 dataset is another widely used dataset benchmark and contains 60,000 color images categorized into 10 distinct classes \cite{Krizhevsky09learningmultiple}.
Each image has a resolution of \(3 \times 32 \times 32\), corresponding to RGB color channels. Compared to MNIST, CIFAR-10 is more challenging due to the diversity and complexity of the objects it represents. 
In our evaluation, we used 150 images from the validation set of CIFAR-10 to evaluate the  ResNet-19 models. 
Just like N-MNIST, the CIFAR10-DVS dataset is a neuromorphic version of the standard CIFAR-10 dataset, adapted for event-based processing in SNNs~\cite{cifar10dvs}. It consists of recordings from a Dynamic Vision Sensor (DVS) that captures changes in pixel intensity over time, resulting in sparse, asynchronous spike events. Each of the original CIFAR-10 images is converted into a short spatiotemporal event stream that reflects temporal dynamics similar to those observed in biological vision. The event streams are stored in AER format, where each event encodes the pixel location, timestamp, and polarity of the change.
To prepare this dataset for efficient encrypted SNN inference, we converted each CIFAR10-DVS sample into 5 time steps, with each time step represented as a frame of dimensions \(2 \times 32 \times 32\). This transformation enables the data to be efficiently fitted into SIMD ciphertexts under the CKKS scheme. 
In comparison to the common transformation approach used in SNN literature, which involves 10 time steps with frame dimensions of \(2 \times 48 \times 48\), our reduced representation results in only a modest accuracy drop of approximately 5 percent in the Resnet-19 models from 73.6\% to 68.10. 
We believe this is an acceptable trade-off given the significant reduction in computational overhead that would have been incurred to process models with inputs of size \(2 \times 48 \times 48\) under HE. 
We evaluated our methods on 150 encrypted preprocessed samples from the validation set of CIFAR10-DVS.

\subsection{HE Security Paramaters}

We pre-determined the CKKS encryption parameter set to enable the most efficient computations within each neural network architecture.
These parameters include the polynomial degree for the CKKS encryption, the number of slots available in each ciphertext, the multiplicative depth, and the scaling modulus of bootstrapping. 
Selecting appropriate encryption parameters is critical, as they directly impact the number of ciphertext slots and the computation depth that can be achieved before and after bootstrapping. However, there is a trade-off as larger encryption parameters result in slower performance, increased key sizes, and higher memory requirements for ciphertext evaluations. Balancing these factors is essential to optimize both efficiency and security. 
Table \ref{tab:fhe_params} shows the parameter sets we used for the different architectures. We selected these parameters to serve as a general configuration for all models within the same architecture thus giving us a matrix for comparative analysis.

\begin{table}[ht]
    \small
    \centering
    \caption{FHE parameter sets for encrypted inference.}
    \label{tab:fhe_params}
    \renewcommand{\arraystretch}{1.2}
    \setlength{\tabcolsep}{6pt} 
    \begin{tabularx}{\linewidth}{|X|r|r|}
        \hline
        \textbf{Parameter} & \textbf{LeNet-5} & \textbf{ResNet-19} \\
        \hline
        Polynomial Degree      & 16,384  & 32,768  \\
        \hline
        Number of SIMD Slots   & 8,192   & 16,384  \\
        \hline
        Multiplicative Depth & 12      & 12      \\
        \hline
        Scaling Factor         & 56      & 56      \\
        \hline
    \end{tabularx}
\end{table}

\section{Results}
\label{sec:results}

We measure the accuracy, latency, and memory used by models built on \SecureSpike. 
Table~\ref{tab:securespike_results} compares the baseline accuracies of models with their encrypted accuracies and also shows the latency and memory profiles of all our models. 

\begin{table}[ht]
    \scriptsize
    \centering
    \caption{Accuracy, Latency, and Memory Usage of Models Using Approximation and Scheme-Switch LIF Neurons Compared to Baseline Plaintext SNNs}
    \label{tab:securespike_results}
    \renewcommand{\arraystretch}{1.1}
    \setlength{\tabcolsep}{3pt} 
    \begin{tabularx}{\linewidth}{|X|p{1.5cm}|p{0.5cm}|p{1.5cm}|p{1cm}|p{1cm}|p{1cm}|}
        \hline
        \textbf{Arch.} & \textbf{Dataset} & \textbf{TS} & \textbf{LIF Type} & \textbf{Acc. (\%)} & \textbf{Lat. (s)} & \textbf{Mem. (GB)} \\
        \hline
        \multirow{6}{*}{LeNet-5} 
          & \multirow{3}{*}{MNIST}    & \multirow{3}{*}{2} 
            & Baseline       & 98.90 & --   & -- \\
        \cline{4-7}
          &                          & 
            & Approx.   & 95.70 & 28  & 4.3 \\
        \cline{4-7}
          &                          & 
            & Scheme-Sw.   & 98.10 & 110   & 9.4  \\
        \cline{2-7}
          & \multirow{3}{*}{N-MNIST} & \multirow{3}{*}{5} 
            & Baseline        & 99.02 & --   & -- \\
        \cline{4-7}
          &                          & 
            & Approx.   & 95.3 & 212  & 11.4  \\
        \cline{4-7}
          &                          & 
            & Scheme-Sw.   & 97.3 & 714 & 25.4 \\
        \hline
        \multirow{6}{*}{ResNet-19} 
          & \multirow{3}{*}{CIFAR-10} & \multirow{3}{*}{2} 
            & Baseline        & 83.19 & --   & -- \\
        \cline{4-7}
          &                           & 
            & Approx.   & 76.0 & 784 & 18.7  \\
        \cline{4-7}
          &                           & 
            & Scheme-Sw.   & 79.3 & 3264 & 85.9 \\
        \cline{2-7}
          & \multirow{3}{*}{\makecell{CIFAR-10 \\ DVS}} & \multirow{3}{*}{5}
            & Baseline        & 68.10 & --   & -- \\
        \cline{4-7}
          &                              & 
            & Approx.   & 64.71 & 1846 & 21.2 \\
        \cline{4-7}
          &                              & 
            & Scheme-Sw.   & 66.00 & 8167 & 93.7 \\
        \hline
    \end{tabularx}
\end{table}

\subsection{Accuracy}

On the  SNN LeNet-5 model using the MNIST dataset, the plaintext model achieved an accuracy of 98.9\%. The encrypted counterpart using the LIF approximation achieved a 95.70\%, reflecting a 3.2\% accuracy drop. The version using scheme switching for LIF evaluation achieved a 98.10\% accuracy, reducing the performance gap to only 0.8\% relative to the plaintext baseline. These results indicate that both methods maintain high accuracy in the encrypted domain, with the scheme-switch approach almost matching the plaintext model. 
On the N-MNIST dataset, the plaintext SNN LeNet-5 model achieved an accuracy of 99.02\%. The encrypted model using LIF approximation achieves 95.3\%, while the scheme-switching approach also matches the plaintext accuracy at 97.3\% further supporting the findings we got from the MNIST dataset.

Using the ResNet-19 models on both the standard CIFAR-10  and the  CIFAR-10 DVS datasets.
The plaintext ResNet-19 model achieved an accuracy of 83.19\% on the CIFAR-10 dataset. The encrypted variant using the LIF approximation attained a 76.0\%. In contrast, the scheme-switching LIF approach improves accuracy in the encrypted domain to 79.3\%, reducing the performance gap to 3.89\%.
On the CIFAR-10 DVS dataset, the plaintext model achieved an accuracy of 68.10\% while the encrypted variant using LIF approximation obtained 64.71\%. 
The ResNet-19 model using the scheme-switching reached a 66.0\% accuracy.

\subsection{Memory Usage }

We compared the memory consumption of the privacy-preserving LeNet-5 and ResNet-19 models by measuring the memory used per inference image. 
Models that utilize scheme switching for LIF evaluation showed significantly higher memory requirements compared to those using the polynomial approximation approach.
Our results also indicated very minimal additional memory overhead across time steps, demonstrating efficient temporal scaling of encrypted inference. This efficiency is primarily due to the rotation reuse strategy employed in \SecureSpike, where the most memory-intensive components (the rotation keys) remain unchanged regardless of the number of time steps. As a result, memory usage stays relatively constant over time.

\subsection{Latency}
Although the scheme-switching evaluation of LIF results in higher-precision models, the latency requirements also significantly increase compared to models that use the approximation approach. This is understandable, as the evaluation of the LIF under TFHE requires unique keys for switching schemes, as well as slower computation. These results illustrate the trade-off one must consider when selecting an appropriate LIF evaluation approach.

\subsection{Comparative Studies compared with Related Works}
We compare the results of our work with those presented in Farzad et al. \cite{farzad} and FHE-DiCNN \cite{Li2023EfficientPC}. 
Table \ref{tab:comparison_results} highlights the superior efficiency and performance of \SecureSpike in encrypted inference. 
Our baseline SNN LeNet-5 model, which uses the approximation approach for LIF, achieved 95.70\% accuracy on MNIST using only two time steps, with an inference time of 28 seconds per image. This represents a substantial reduction in computational and latency requirements compared to the work by Farzad et al. \cite{farzad}, which required 40 time steps and 930 seconds to evaluate a single image from the Fashion-MNIST dataset using the same model architecture.
Similarly, when compared to FHE-DiCNN \cite{Li2023EfficientPC}, our SNN LeNet-5 model with LIF approximation computes 60,000 neurons on a consumer-grade CPU in just 28 seconds. In contrast, FHE-DiCNN requires approximately 1,500 seconds on a high-performance CPU to perform the same computation.
For CIFAR-10, our ResNet-19 model achieved 76.0\% using the approximation of the LIF in only two time steps, with a latency of 784 seconds. While no existing works utilize this complex architecture or dataset, Farzad et al. \cite{farzad} estimate an inference latency of 901,800 seconds on a secure SNN AlexNet model with 60 time steps, on Fashion MNIST. This performance and capability difference between \SecureSpike and related works shows our distinguished benefits. 

\begin{table}[ht]
    \scriptsize
    \centering
    \caption{Comparison of \SecureSpike LeNet-5 Model to Related Works}
    \label{tab:comparison_results}
    \setlength{\tabcolsep}{3pt} 
    \renewcommand{\arraystretch}{1.1}
    \begin{tabularx}{\linewidth}{|X|c|c|c|c|}
        \hline
        \textbf{Model} & \textbf{Data} & \textbf{TS} & \textbf{Latency} & \textbf{Acc. (\%)} \\
        \hline
        \textbf{\SecureSpike LeNet-5} & \textbf{MNIST} & \textbf{2} & \textbf{28} & \textbf{95.7} \\
        \hline
        \textbf{\SecureSpike ResNet-19} & \textbf{CIFAR-10} & \textbf{2} & \textbf{784} & \textbf{76.0} \\
        \hline
        Farzad LeNet-5~\cite{farzad} & F-MNIST & 40 & 930 & 96.5 \\
        \hline
        Farzad AlexNet~\cite{farzad} & F-MNIST & 60 & 901,800 & -- \\
        \hline
        FHE-DiCNN~\cite{Li2023EfficientPC} & MNIST & 1 & 1,500 & 95.2 \\
        \hline
    \end{tabularx}
\end{table}

\section{Conclusion}
\label{sec:conclusion}

In this work, we introduced \SecureSpike, an open-source novel framework for privacy-preserving inference in SNNs using the CKKS HE scheme. To enable efficient and high-precision evaluation of the LIF activation function under HE constraints, we proposed and implement two novel HE-friendly LIF algorithms.
Built on top of \SecureSpike, we implemented the LeNet-5 and ResNet-19 SNN architectures also proposing multiple architectural optimizations for HE-friendly SNNs. 
We infer the models on encrypted data across MNIST, Neuromorphic MNIST, CIFAR-10, and CIFAR10-DVS datasets.
Our findings shows the potential and application of SNNs in privacy-preserving machine learning, especially in energy-constrained environments and event-driven data processing scenarios.

Unlike prior works, which are constrained to shallow SNNs and small-scale datasets, \SecureSpike allows for the development of deep privacy-preserving SNN models. The models built on \SecureSpike consistently delivers superior accuracy while achieving significantly lower inference latency compared to prior works. 
On the LeNet-5 architecture, our LIF approximation-based model outperforms the work of Farzad et al.~\cite{farzad} by approximately 34$\times$, and FHE-DiCNN~\cite{Li2023EfficientPC} by approximately 50$\times$ in terms of inference speed while also showing better accuracies in both cases.
Future work will focus on exploring end-to-end encrypted training and inference of SNNs, as well as developing more architecture-level optimizations to further reduce memory overhead and improve inference latency, particularly in the scheme-switching approach to evaluating LIF.

\bibliographystyle{IEEEtran}
\bibliography{paper}

\begin{thebibliography}{10}
\providecommand{\url}[1]{#1}
\csname url@samestyle\endcsname
\providecommand{\newblock}{\relax}
\providecommand{\bibinfo}[2]{#2}
\providecommand{\BIBentrySTDinterwordspacing}{\spaceskip=0pt\relax}
\providecommand{\BIBentryALTinterwordstretchfactor}{4}
\providecommand{\BIBentryALTinterwordspacing}{\spaceskip=\fontdimen2\font plus
\BIBentryALTinterwordstretchfactor\fontdimen3\font minus \fontdimen4\font\relax}
\providecommand{\BIBforeignlanguage}[2]{{%
\expandafter\ifx\csname l@#1\endcsname\relax
\typeout{** WARNING: IEEEtran.bst: No hyphenation pattern has been}%
\typeout{** loaded for the language `#1'. Using the pattern for}%
\typeout{** the default language instead.}%
\else
\language=\csname l@#1\endcsname
\fi
#2}}
\providecommand{\BIBdecl}{\relax}
\BIBdecl

\bibitem{ml_applications}
\BIBentryALTinterwordspacing
N.~Duggal, ``Top 10 machine learning applications and examples in 2024,'' Sep 3, 2024. [Online]. Available: \url{https://www.simplilearn.com/tutorials/machine-learning-tutorial/machine-learning-applications}
\BIBentrySTDinterwordspacing

\bibitem{khan_sok}
U.~Z. . A. S.~Q. Asifullah~Khan, Anabia~Sohail, ``A survey of the recent architectures of deep convolutional neural networks,'' \emph{Artificial Intelligence Review}, 2020.

\bibitem{patrick_fpga}
P.~Plagwitz, F.~Hannig, J.~Teich, and O.~Keszocze, ``Snn vs. cnn implementations on fpgas: An empirical evaluation,'' in \emph{Applied Reconfigurable Computing. Architectures, Tools, and Applications}, I.~Skliarova, P.~Brox~Jim{\'e}nez, M.~V{\'e}stias, and P.~C. Diniz, Eds.\hskip 1em plus 0.5em minus 0.4em\relax Cham: Springer Nature Switzerland, 2024, pp. 3--18.

\bibitem{ClaudioCimarelli-2025}
\BIBentryALTinterwordspacing
C.~Cimarelli, J.~A. Millan-Romera, H.~Voos, and J.~L. Sanchez-Lopez, ``Hardware, algorithms, and applications of the neuromorphic vision sensor: a review,'' 2025. [Online]. Available: \url{https://arxiv.org/abs/2504.08588}
\BIBentrySTDinterwordspacing

\bibitem{GregorLenz-2024}
\BIBentryALTinterwordspacing
G.~Lenz, G.~Orchard, and S.~Sheik, ``Ultra-low-power image classification on neuromorphic hardware,'' 2024. [Online]. Available: \url{https://arxiv.org/abs/2309.16795}
\BIBentrySTDinterwordspacing

\bibitem{IntelLoihi2}
Intel, ``Taking neuromorphic computing to the next level with loihi 2 technology brief,'' \url{https://www.intel.com/content/www/us/en/research/neuromorphic-computing-loihi-2-technology-brief.html}, 2021, accessed: 03-19-2024.

\bibitem{xu}
M.~Xu, X.~Chen, A.~Sun, X.~Zhang, and X.~Chen, ``A novel event-driven spiking convolutional neural network for electromyography pattern recognition,'' \emph{IEEE Transactions on Biomedical Engineering}, vol.~70, no.~9, pp. 2604--2615, 2023.

\bibitem{xu2023novel}
------, ``A novel event-driven spiking convolutional neural network for electromyography pattern recognition,'' \emph{IEEE Transactions on Biomedical Engineering}, vol.~70, no.~9, pp. 2604--2615, 2023.

\bibitem{9936637}
R.~Podschwadt, D.~Takabi, P.~Hu, M.~H. Rafiei, and Z.~Cai, ``A survey of deep learning architectures for privacy-preserving machine learning with fully homomorphic encryption,'' \emph{IEEE Access}, vol.~10, pp. 117\,477--117\,500, 2022.

\bibitem{zhang}
Q.~Zhang, C.~Xin, and H.~Wu, ``Privacy-preserving deep learning based on multiparty secure computation: A survey,'' \emph{IEEE Internet of Things Journal}, vol.~8, no.~13, pp. 10\,412--10\,429, 2021.

\bibitem{Gong}
M.~Gong, Y.~Xie, K.~Pan, K.~Feng, and A.~Qin, ``A survey on differentially private machine learning [review article],'' \emph{IEEE Computational Intelligence Magazine}, vol.~15, no.~2, pp. 49--64, 2020.

\bibitem{Li2023ASO}
\BIBentryALTinterwordspacing
X.~Li, B.~Zhao, G.~Yang, T.~Xiang, J.~Weng, and R.~H. Deng, ``A survey of secure computation using trusted execution environments,'' \emph{ArXiv}, vol. abs/2302.12150, 2023. [Online]. Available: \url{https://api.semanticscholar.org/CorpusID:257102679}
\BIBentrySTDinterwordspacing

\bibitem{njungle2025guardianml}
N.~B. Njungle, E.~Jahns, Z.~Wu, L.~Mastromauro, M.~Stojkov, and M.~Kinsy, ``Guardianml: Anatomy of privacy-preserving machine learning techniques and frameworks,'' \emph{IEEE Access}, 2025.

\bibitem{ckks}
\BIBentryALTinterwordspacing
J.~H. Cheon, A.~Kim, M.~Kim, and Y.~Song, ``Homomorphic encryption for arithmetic of approximate numbers,'' Cryptology {ePrint} Archive, Paper 2016/421, 2016. [Online]. Available: \url{https://eprint.iacr.org/2016/421}
\BIBentrySTDinterwordspacing

\bibitem{farzad}
\BIBentryALTinterwordspacing
F.~Nikfam, R.~Casaburi, A.~Marchisio, M.~Martina, and M.~Shafique, ``A homomorphic encryption framework for privacy-preserving spiking neural networks,'' \emph{Information}, vol.~14, no.~10, 2023. [Online]. Available: \url{https://www.mdpi.com/2078-2489/14/10/537}
\BIBentrySTDinterwordspacing

\bibitem{tfhe}
\BIBentryALTinterwordspacing
I.~Chillotti, N.~Gama, M.~Georgieva, and M.~Izabachène, ``Tfhe: Fast fully homomorphic encryption over the torus,'' Cryptology ePrint Archive, Paper 2018/421, 2018, \url{https://eprint.iacr.org/2018/421}. [Online]. Available: \url{https://eprint.iacr.org/2018/421}
\BIBentrySTDinterwordspacing

\bibitem{farzad_thesis}
R.~Casaburi, ``Homomorphic encryption for spiking neural networks,'' Ph.D. dissertation, POLITECNICO DI TORINO, 2022.

\bibitem{bfv}
\BIBentryALTinterwordspacing
J.~Fan and F.~Vercauteren, ``Somewhat practical fully homomorphic encryption,'' \emph{IACR Cryptol. ePrint Arch.}, vol. 2012, p. 144, 2012. [Online]. Available: \url{https://api.semanticscholar.org/CorpusID:1467571}
\BIBentrySTDinterwordspacing

\bibitem{Li2023EfficientPC}
\BIBentryALTinterwordspacing
P.~Li, H.~Huang, T.~Gao, J.~Guo, and J.~Duan, ``Efficient privacy-preserving convolutional spiking neural networks with fhe,'' \emph{ArXiv}, vol. abs/2309.09025, 2023. [Online]. Available: \url{https://api.semanticscholar.org/CorpusID:262046057}
\BIBentrySTDinterwordspacing

\bibitem{gentry}
C.~Gentry, ``A fully homomorphic encryption scheme,'' Ph.D. dissertation, Stanford University, 2009, \url{crypto.stanford.edu/craig}.

\bibitem{fhe_vul}
\BIBentryALTinterwordspacing
C.~Marcolla, V.~Sucasas, M.~Manzano, R.~Bassoli, F.~H. Fitzek, and N.~Aaraj, ``Survey on fully homomorphic encryption, theory, and applications,'' Cryptology ePrint Archive, Paper 2022/1602, 2022, \url{https://eprint.iacr.org/2022/1602}. [Online]. Available: \url{https://eprint.iacr.org/2022/1602}
\BIBentrySTDinterwordspacing

\bibitem{bgv}
N.~Aggarwal, C.~Gupta, and I.~Sharma, ``Fully homomorphic symmetric scheme without bootstrapping,'' pp. 14--17, 2014.

\bibitem{ringlwe}
\BIBentryALTinterwordspacing
V.~Lyubashevsky, C.~Peikert, and O.~Regev, ``On ideal lattices and learning with errors over rings,'' Cryptology {ePrint} Archive, Paper 2012/230, 2012. [Online]. Available: \url{https://eprint.iacr.org/2012/230}
\BIBentrySTDinterwordspacing

\bibitem{rns_ckks}
\BIBentryALTinterwordspacing
J.~H. Cheon, K.~Han, A.~Kim, M.~Kim, and Y.~Song, ``A full {RNS} variant of approximate homomorphic encryption,'' Cryptology {ePrint} Archive, Paper 2018/931, 2018. [Online]. Available: \url{https://eprint.iacr.org/2018/931}
\BIBentrySTDinterwordspacing

\bibitem{sealcrypto}
``{M}icrosoft {SEAL} (release 4.1),'' \url{https://github.com/Microsoft/SEAL}, Jan. 2023, microsoft Research, Redmond, WA., Accessed: 2024-02-27.

\bibitem{helib}
\BIBentryALTinterwordspacing
S.~Halevi and V.~Shoup, ``Design and implementation of helib: a homomorphic encryption library,'' Cryptology ePrint Archive, Paper 2020/1481, 2020, \url{https://eprint.iacr.org/2020/1481}. [Online]. Available: \url{https://eprint.iacr.org/2020/1481}
\BIBentrySTDinterwordspacing

\bibitem{TFHErs}
Zama, ``{TFHE-rs: A Pure Rust Implementation of the TFHE Scheme for Boolean and Integer Arithmetics Over Encrypted Data},'' 2022, \url{https://github.com/zama-ai/tfhe-rs}.

\bibitem{OpenFHE}
\BIBentryALTinterwordspacing
A.~A. Badawi, J.~Bates, F.~Bergamaschi, D.~B. Cousins, S.~Erabelli, N.~Genise, S.~Halevi, H.~Hunt, A.~Kim, Y.~Lee, Z.~Liu, D.~Micciancio, I.~Quah, Y.~Polyakov, S.~R.V., K.~Rohloff, J.~Saylor, D.~Suponitsky, M.~Triplett, V.~Vaikuntanathan, and V.~Zucca, ``Openfhe: Open-source fully homomorphic encryption library,'' Cryptology ePrint Archive, Paper 2022/915, 2022, \url{https://eprint.iacr.org/2022/915}. [Online]. Available: \url{https://eprint.iacr.org/2022/915}
\BIBentrySTDinterwordspacing

\bibitem{njungle2025safety}
N.~B. Njungle, M.~Stojkov, and M.~A. Kinsy, ``A safety-centric analysis and benchmarks of modern open-source homomorphic encryption libraries,'' 2025.

\bibitem{ghosh2009spiking}
S.~Ghosh-Dastidar and H.~Adeli, ``Spiking neural networks,'' \emph{International journal of neural systems}, vol.~19, no.~04, pp. 295--308, 2009.

\bibitem{marchisio2020spiking}
A.~Marchisio, G.~Nanfa, F.~Khalid, M.~A. Hanif, M.~Martina, and M.~Shafique, ``Is spiking secure? a comparative study on the security vulnerabilities of spiking and deep neural networks,'' in \emph{2020 International Joint Conference on Neural Networks (IJCNN)}.\hskip 1em plus 0.5em minus 0.4em\relax IEEE, 2020, pp. 1--8.

\bibitem{spikinghard}
A.~Shrestha, H.~Fang, Z.~Mei, D.~P. Rider, Q.~Wu, and Q.~Qiu, ``A survey on neuromorphic computing: Models and hardware,'' \emph{IEEE Circuits and Systems Magazine}, vol.~22, no.~2, pp. 6--35, 2022.

\bibitem{han2021survey}
C.~S. Han and K.~M. Lee, ``A survey on spiking neural networks,'' \emph{International Journal of Fuzzy Logic and Intelligent Systems}, vol.~21, no.~4, pp. 317--337, 2021.

\bibitem{ho2021tcl}
N.-D. Ho and I.-J. Chang, ``Tcl: an ann-to-snn conversion with trainable clipping layers,'' in \emph{2021 58th ACM/IEEE Design Automation Conference (DAC)}.\hskip 1em plus 0.5em minus 0.4em\relax IEEE, 2021, pp. 793--798.

\bibitem{almomani2019comparative}
A.~Almomani, M.~Alauthman, M.~Alweshah, O.~Dorgham, and F.~Albalas, ``A comparative study on spiking neural network encoding schema: implemented with cloud computing,'' \emph{Cluster Computing}, vol.~22, pp. 419--433, 2019.

\bibitem{yang2023spiking}
Y.~Yang, J.~Ren, and F.~Duan, ``The spiking rates inspired encoder and decoder for spiking neural networks: an illustration of hand gesture recognition,'' \emph{Cognitive Computation}, vol.~15, no.~4, pp. 1257--1272, 2023.

\bibitem{kim2022rate}
Y.~Kim, H.~Park, A.~Moitra, A.~Bhattacharjee, Y.~Venkatesha, and P.~Panda, ``Rate coding or direct coding: Which one is better for accurate, robust, and energy-efficient spiking neural networks?'' in \emph{ICASSP 2022-2022 IEEE International Conference on Acoustics, Speech and Signal Processing (ICASSP)}.\hskip 1em plus 0.5em minus 0.4em\relax IEEE, 2022, pp. 71--75.

\bibitem{mostafa2017supervised}
H.~Mostafa, ``Supervised learning based on temporal coding in spiking neural networks,'' \emph{IEEE transactions on neural networks and learning systems}, vol.~29, no.~7, pp. 3227--3235, 2017.

\bibitem{encoders}
\BIBentryALTinterwordspacing
E.~Forno, V.~Fra, R.~Pignari, E.~Macii, and G.~Urgese, ``Spike encoding techniques for iot time-varying signals benchmarked on a neuromorphic classification task,'' \emph{Frontiers in Neuroscience}, vol.~16, 2022. [Online]. Available: \url{https://www.frontiersin.org/ journals/neuroscience/articles/10.3389/fnins.2022.999029}
\BIBentrySTDinterwordspacing

\bibitem{namatevs2017deep}
I.~Namat{\=e}vs, ``Deep convolutional neural networks: Structure, feature extraction and training,'' \emph{Information Technology and Management Science}, vol.~20, no.~1, pp. 40--47, 2017.

\bibitem{Kamath2019}
U.~Kamath, J.~Liu, and J.~Whitaker, \emph{Convolutional Neural Networks}.\hskip 1em plus 0.5em minus 0.4em\relax Cham: Springer International Publishing, 2019, pp. 263--314.

\bibitem{JIANG2022102954}
\BIBentryALTinterwordspacing
J.~Jiang, D.~Huang, J.~Du, Y.~Lu, and X.~Liao, ``Optimizing small channel 3d convolution on gpu with tensor core,'' \emph{Parallel Computing}, vol. 113, p. 102954, 2022. [Online]. Available: \url{https://www.sciencedirect.com/science/article/pii/S0167819122000473}
\BIBentrySTDinterwordspacing

\bibitem{gazelle}
\BIBentryALTinterwordspacing
C.~Juvekar, V.~Vaikuntanathan, and A.~Chandrakasan, ``{GAZELLE}: A low latency framework for secure neural network inference,'' in \emph{27th USENIX Security Symposium (USENIX Security 18)}.\hskip 1em plus 0.5em minus 0.4em\relax Baltimore, MD: USENIX Association, Aug. 2018, pp. 1651--1669. [Online]. Available: \url{https://www.usenix.org/conference/usenixsecurity18/presentation/juvekar}
\BIBentrySTDinterwordspacing

\bibitem{kim_fhe}
D.~Kim and C.~Guyot, ``Optimized privacy-preserving cnn inference with fully homomorphic encryption,'' \emph{IEEE Transactions on Information Forensics and Security}, vol.~18, pp. 2175--2187, 2023.

\bibitem{lee22e}
\BIBentryALTinterwordspacing
E.~Lee, J.-W. Lee, J.~Lee, Y.-S. Kim, Y.~Kim, J.-S. No, and W.~Choi, ``Low-complexity deep convolutional neural networks on fully homomorphic encryption using multiplexed parallel convolutions,'' in \emph{Proceedings of the 39th International Conference on Machine Learning}, ser. Proceedings of Machine Learning Research, K.~Chaudhuri, S.~Jegelka, L.~Song, C.~Szepesvari, G.~Niu, and S.~Sabato, Eds., vol. 162.\hskip 1em plus 0.5em minus 0.4em\relax PMLR, 17--23 Jul 2022, pp. 12\,403--12\,422. [Online]. Available: \url{https://proceedings.mlr.press/v162/lee22e.html}
\BIBentrySTDinterwordspacing

\bibitem{lowmemory20}
\BIBentryALTinterwordspacing
L.~Rovida and A.~Leporati, ``Encrypted image classification with low memory footprint using fully homomorphic encryption,'' Cryptology {ePrint} Archive, Paper 2024/460, 2024. [Online]. Available: \url{https://eprint.iacr.org/2024/460}
\BIBentrySTDinterwordspacing

\bibitem{gholamalinezhad2020pooling}
\BIBentryALTinterwordspacing
H.~Gholamalinezhad and H.~Khosravi, ``Pooling methods in deep neural networks, a review,'' 2020. [Online]. Available: \url{https://arxiv.org/abs/2009.07485}
\BIBentrySTDinterwordspacing

\bibitem{PASSRICHA20195}
\BIBentryALTinterwordspacing
V.~Passricha and R.~K. Aggarwal, ``Chapter 2 - end-to-end acoustic modeling using convolutional neural networks,'' in \emph{Intelligent Speech Signal Processing}, N.~Dey, Ed.\hskip 1em plus 0.5em minus 0.4em\relax Academic Press, 2019, pp. 5--37. [Online]. Available: \url{https://www.sciencedirect.com/science/article/pii/B9780128181300000027}
\BIBentrySTDinterwordspacing

\bibitem{intro_cnn}
\BIBentryALTinterwordspacing
geeksforgeeks, ``Introduction to convolution neural network,'' 2018. [Online]. Available: \url{https://www.geeksforgeeks.org/introduction-convolution-neural-network/}
\BIBentrySTDinterwordspacing

\bibitem{DUBEY202292}
\BIBentryALTinterwordspacing
S.~R. Dubey, S.~K. Singh, and B.~B. Chaudhuri, ``Activation functions in deep learning: A comprehensive survey and benchmark,'' \emph{Neurocomputing}, vol. 503, pp. 92--108, 2022. [Online]. Available: \url{https://www.sciencedirect.com/science/article/pii/S0925231222008426}
\BIBentrySTDinterwordspacing

\bibitem{nunes2022spiking}
J.~D. Nunes, M.~Carvalho, D.~Carneiro, and J.~S. Cardoso, ``Spiking neural networks: A survey,'' \emph{IEEE Access}, vol.~10, pp. 60\,738--60\,764, 2022.

\bibitem{teeter2018generalized}
C.~Teeter, R.~Iyer, V.~Menon, N.~Gouwens, D.~Feng, J.~Berg, A.~Szafer, N.~Cain, H.~Zeng, M.~Hawrylycz \emph{et~al.}, ``Generalized leaky integrate-and-fire models classify multiple neuron types,'' \emph{Nature communications}, vol.~9, no.~1, p. 709, 2018.

\bibitem{gil2012non}
L.~A. Gil-Alana and J.~C. Cuestas, ``A non-linear approach with long range dependence based on chebyshev polynomials,'' 2012.

\bibitem{fhew}
\BIBentryALTinterwordspacing
I.~Chillotti, N.~Gama, M.~Georgieva, and M.~Izabachène, ``Faster fully homomorphic encryption: Bootstrapping in less than 0.1 seconds,'' Cryptology {ePrint} Archive, Paper 2016/870, 2016. [Online]. Available: \url{https://eprint.iacr.org/2016/870}
\BIBentrySTDinterwordspacing

\bibitem{chimera}
\BIBentryALTinterwordspacing
C.~Boura, N.~Gama, M.~Georgieva, and D.~Jetchev, ``{CHIMERA}: Combining ring-{LWE}-based fully homomorphic encryption schemes,'' Cryptology {ePrint} Archive, Paper 2018/758, 2018. [Online]. Available: \url{https://eprint.iacr.org/2018/758}
\BIBentrySTDinterwordspacing

\bibitem{deng2012mnist}
L.~Deng, ``The mnist database of handwritten digit images for machine learning research,'' \emph{IEEE Signal Processing Magazine}, vol.~29, no.~6, pp. 141--142, 2012.

\bibitem{nmnist}
\BIBentryALTinterwordspacing
G.~Orchard, A.~Jayawant, G.~K. Cohen, and N.~Thakor, ``Converting static image datasets to spiking neuromorphic datasets using saccades,'' \emph{Frontiers in Neuroscience}, vol.~9, 2015. [Online]. Available: \url{https://www.frontiersin.org/journals/neuroscience/articles/10.3389/fnins.2015.00437}
\BIBentrySTDinterwordspacing

\bibitem{Krizhevsky09learningmultiple}
A.~Krizhevsky, ``Learning multiple layers of features from tiny images,'' Tech. Rep., 2009.

\bibitem{cifar10dvs}
\BIBentryALTinterwordspacing
H.~Li, H.~Liu, X.~Ji, G.~Li, and L.~Shi, ``Cifar10-dvs: An event-stream dataset for object classification,'' \emph{Frontiers in Neuroscience}, vol. Volume 11 - 2017, 2017. [Online]. Available: \url{https://www.frontiersin.org/journals/neuroscience/articles/10.3389/fnins.2017.00309}
\BIBentrySTDinterwordspacing

\end{thebibliography}


\section{Appendix}

\subsection{Chebyshev polynomial Approximation}

Chebyshev polynomials  are denoted as $T_n(x)$, and defined recursively as:
\begin{align}
    T_0(x) &= 1 \\
    T_1(x) &= x \\
    T_n(x) &= 2x T_{n-1}(x) - T_{n-2}(x), \quad \text{for } n \geq 2
\end{align}

These polynomials exhibit several important properties. Mainly, a polynomial of degree $n$ has exactly $n$ roots in the interval $[-1, 1]$, which are distributed as:
\begin{align}
x_k = \cos\left(\frac{k\pi}{n}\right), \quad \text{where } 0 \leq k < n.
\end{align}

This distribution of roots ensures that Chebyshev polynomials achieve the best uniform approximation to a smooth function over $[-1, 1]$. This makes them highly efficient for approximating non-linear functions.
The approximation of a non-linear function $f(x)$ over the interval $[-1, 1]$ can be achieved by expressing it as a  series of Chebyshev polynomials:
\begin{align}
f(x) \approx \sum_{n=0}^{N} c_n T_n(x),
\end{align}
where $c_n$ are the coefficients determined by the function $f(x)$ and $N$ is the degree of the polynomial.

\subsection{Spiking Neural Network Visual Representations}

\begin{figure}[http]
    \begin{center}
    \includegraphics[width=\linewidth]{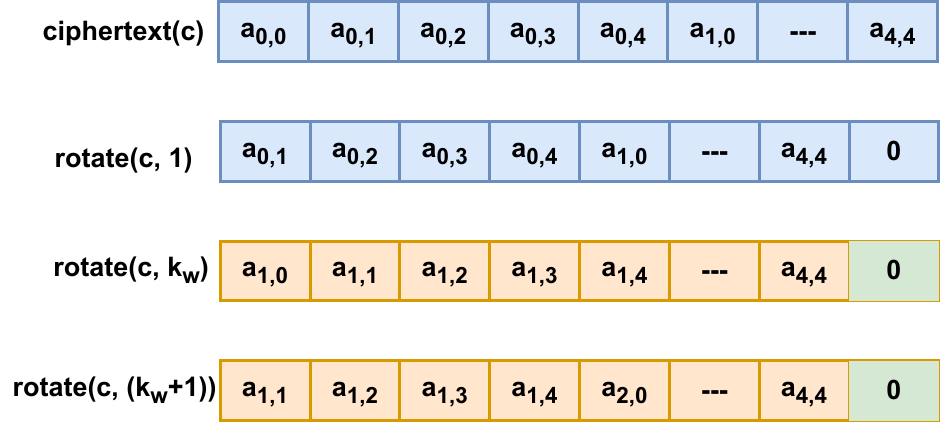}
    \captionsetup{justification=centering}
    \caption{Rotated Ciphertexts for convolution. The number of rotations is a total of $k^2-1$ where $k$ is the kernel width.}
    \label{fig:con_rots}
    \end{center}

\end{figure}

\begin{figure}[http]
    \begin{center}
    \includegraphics[width=\linewidth]{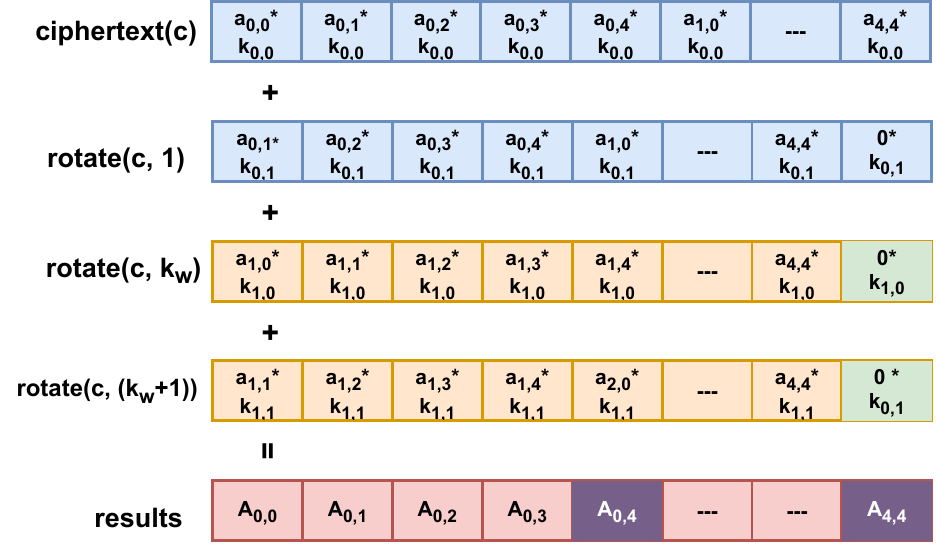}
    \captionsetup{justification=centering}
    \caption{Vector Encoding Convolution: The repeated kernel values are multiplied with equivalent rotated ciphertexts and summed to produce results.}
    \label{fig:con_sum}
    \end{center}
\end{figure}

\begin{figure}[http]
    \begin{center}
    \includegraphics[width=\linewidth]{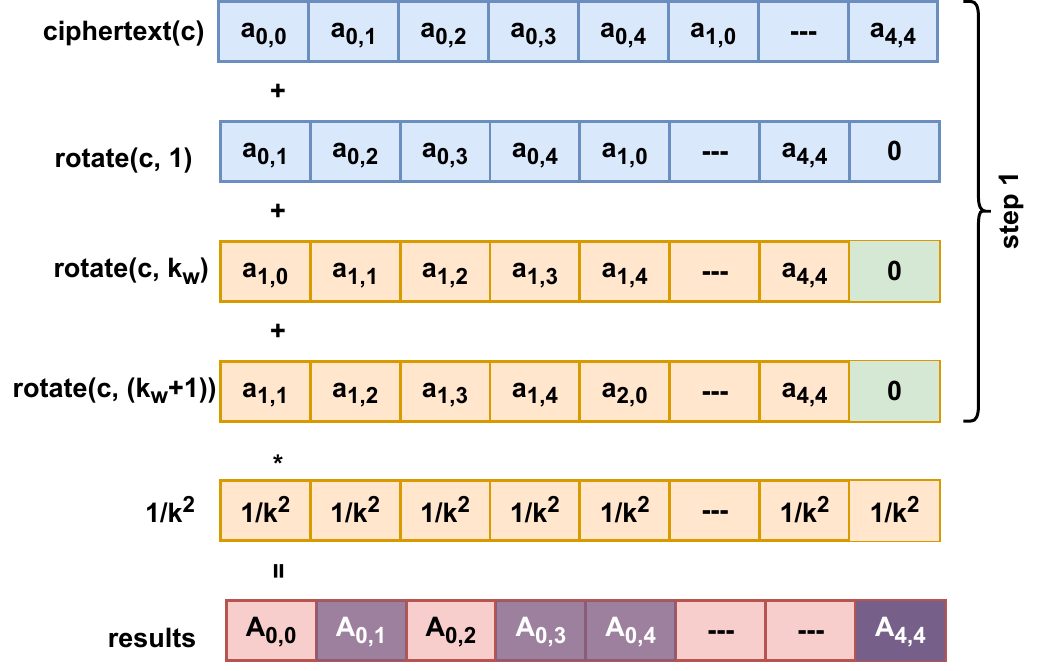}
    \captionsetup{justification=centering}
    \caption{Vector Encoding Secure Average Pooling: These rotations are then summed and multiplied by $1/k^2$}
    \label{fig:pooling}
    \end{center}
\end{figure}

\subsection{Model Architectures}
\begin{table}[ht]
\small
\centering
\caption{Layer configurations for LeNet-5 and ResNet-19. Conv: Convolution, FC: Fully Connected, RB: Residual Block. Config shows (Kernel Size, Padding, Stride).}
\label{tab:arch_com}
\renewcommand{\arraystretch}{1.2}
\setlength{\tabcolsep}{4pt}
\begin{tabularx}{\linewidth}{|X|l|r|r|X|}
\hline
\textbf{Architecture} & \textbf{Layer} & \textbf{In Ch.} & \textbf{Out Ch.} & \textbf{Kernel, Padding, Stride} \\ 
\hline

\multirow{7}{*}{LeNet-5}        
    & Conv + LIF & 1 (or 2) & 6   & $5\times5$, 1, 0 \\ \cline{2-5}
    & AvgPool    & 6        & 6   & $2\times2$, 2, 0 \\ \cline{2-5}
    & Conv + LIF & 6        & 16  & $5\times5$, 1, 0 \\ \cline{2-5}
    & AvgPool    & 16       & 16  & $2\times2$, 2, 0 \\ \cline{2-5}
    & FC + LIF   & \makecell{256 \\ (or 576)} & 120 & -- \\ \cline{2-5}
    & FC + LIF   & 120      & 84  & -- \\ \cline{2-5}
    & FC         & 84       & 10  & -- \\ 
\hline

\multirow{6}{*}{ResNet-19}        
    & Conv + LIF & 3 (or 2) & 16  & $3\times3$, 1, 1 \\ \cline{2-5}
    & 3 RBs      & 16       & 16  & \makecell{$3\times3$, 1 \\ (1,1,1)} \\ \cline{2-5}
    & 3 RBs      & 16       & 32  & \makecell{$3\times3$, 1 \\ (2,1,1)} \\ \cline{2-5}
    & 2 RBs      & 32       & 64  & \makecell{$3\times3$, 1 \\ (2,1)}  \\ \cline{2-5}
    & AvgPool    & 64       & 64  & $8\times8$, 1, 0 \\ \cline{2-5}
    & FC         & 64       & 10  & -- \\ 
\hline
\end{tabularx}
\end{table}

\end{document}